# VOLUMETRIC AND VISCOSIMETRIC MEASUREMENTS FOR METHANOL + CH$_3$-O-(CH$_2$CH$_2$O)$_n$-CH$_3$ ($n$ = 2,3,4) MIXTURES AT (293.15-303.15) K AND ATMOSPHERIC PRESSURE. APPLICATION OF THE ERAS MODEL

**Juan A. González,\* Francisco J. Martínez, Luis F. Sanz, Fernando Hevia, Isaías García de la Fuente, José C. Cobos**

G.E.T.E.F., Departamento de Física Aplicada, Facultad de Ciencias, Universidad de Valladolid. Paseo de Belén, 7, 47011 Valladolid, Spain.

\*e-mail: jagl@termo.uva.es; Tel: +34-983-423757

**ABSTRACT**

Densities, $\rho$, and kinematic viscosities, $\nu$, have been determined at atmospheric pressure and at (293.15-303.15) K for binary mixtures formed by methanol and one linear polyether of the type $CH_3$-O-$(CH_2CH_2O)_n$-$CH_3$ ($n$ =2,3,4). Measurements on $\rho$ and $\nu$ were carried out, respectively, using an Anton Paar DMA 602 vibrating-tube densimeter and an Ubbelohde viscosimeter. The $\rho$ values are used to compute excess molar volumes, $V_m^E$, and, together with $\nu$ results, dynamic viscosities ($\eta$). Deviations from linear dependence on mole fraction for viscosity, $\Delta\eta$, are also provided. Different semi-empirical equations have been employed to correlate viscosity data. Particularly, the equations used are: Grunberg-Nissan, Hind, Frenkel, Katti-Chaudhri, McAllister and Heric. Calculations show that better results are obtained from the Hind equation. The $V_m^E$ values are large and negative and contrast with the positive excess molar enthalpies, $H_m^E$, available in the literature, for these systems. This indicates that structural effects are dominant. The $\Delta\eta$ results are positive and correlate well with the difference in volume of the mixture compounds confirming the importance of structural effects. The temperature dependences of $\eta$ and of the molar volume have been used to calculate enthalpies, entropies and Gibbs energies, $\Delta G^*$, of viscous flow. It is demonstrated that $\Delta G^*$ is essentially determined by enthalpic effects. Methanol + $CH_3$-O-$(CH_2CH_2O)_n$-$CH_3$ mixtures have been treated in the framework of the ERAS model. Results on $H_m^E$ are acceptable, while the composition dependence of the $V_m^E$ curves is poorly represented. This has been ascribed to the existence of strong dipolar and structural effects in the present solutions.

1.  **INTRODUCTION**

Alcohol + ether mixtures show a large variety of applications. For example, it is well-known that, due to their octane-enhancing and pollution reducing properties, these systems are common gasoline additives [1,2]. Mixtures formed by an alcohol as refrigerant and an absorbent (polyethers or polyethylene glycols, PEG) are good working fluids for absorption refrigerant machines since their use leads to an improvement of the cycle machine [3]. Alkanol + ether systems are also relevant in the industry since alkanols are basic components in the synthesis of oxaalkanes and therefore are contained as an impurity. The complex water + PEG systems, widely used in biochemical and biomedical processes, can be tentatively modelled by means of mixtures of short chain 1-alkanols with linear polyethers [4]. Mixtures with cyclic ethers have also progressively gained attraction in biotechnology [5,6]

This work is a new contribution to the general experimental and theoretical investigation on 1-alkanol + linear or cyclic ether systems which has been undertaken by our research group over the past decades. Thus, we have provided excess molar volumes, $V_m^E$, speeds of sound or adiabatic compressibilities [7-14], excess molar enthalpies, $H_m^E$, [15-18] or permittivities and refractive indices [19,20] for this type of mixtures. In addition, we have conducted detailed studies on 1-alkanol + ether systems [7,14,21-23] using a set of different models such as DISQUAC [24], ERAS [25], Flory [26] or the Kirkwood-Buff formalism [27]. As continuation, we provide now densities ($\rho$), kinematic ($\nu$) and dynamic ($\eta$) viscosities, $V_m^E$ and deviations of $\eta$ from the linear dependence on mole fraction for the systems methanol + 2,5,8-trioxanonane, or + 2,5,8,11-tetraoxadodecane, or + 2,5,8,11,14-pentaoxapentadecane at (293.15, 298.15, 303.15) K. Some experimental results on these systems are already available in the literature: $V_m^E$ for the mentioned mixtures at 298.15 K [11,28]; and $V_m^E$ and $\eta$ at 293.15 K and 303.15 K for the methanol + 2,5,8,11,14-pentaoxapentadecane mixture [29,30]. Viscosity data have been correlated by means of a number of semi-empirical equations. For dynamic viscosities the equations applied are: Grunberg-Nissan [31], Hind [32], Frenkel [33], Katti-Chaudhri [34]. For kinematic viscosities, the equations used are those developed by McAllister [35] and by Heric [36]. Finally, we have also applied the ERAS model to the investigated mixtures in order to explore the ability of the theory to describe simultaneously $H_m^E$ and $V_m^E$ results for mixtures where strong dipolar interactions are expected. In this regards, we remark that $CH_3$-O-$(CH_2CH_2O)_n$-$CH_3$ ($n = 3,4$) + $n$-alkane mixtures show miscibility gaps at temperatures not far from 298.15 K. For example, the upper critical solution temperatures of the mixtures 2,5,8,11-tetraoxadodecane + $n$-$C_{12}$ or of 2,5,8,11,14-pentaoxapentadecane + $n$-$C_8$ are, respectively, 280.8 K [37] and 281.8 K [38], which points out to the existence of strong dipolar interactions between ether molecules. From our point of view, this makes that the ERAS treatment of 1-propanol + 2,5,8-trioxanonane, or + 2,5,8,11-tetraoxadodecane systems assuming that these ethers are self-associated species [39], is not consistent.

## 2. EXPERIMENTAL

*2.1 Materials*

All the compounds were used as received, without further purification. However, prior to measurements, they were carefully dried with molecular sieve (Union Carbide, Type 3A) added to the liquids. Table 1 shows some basic information of the pure chemicals employed such as their source and purity. On the other hand, the $\rho$, and $\eta$ values measured in this work for the pure compounds are in good agreement with literature data (Table 2).

*2.2 Apparatus and procedures*

Binary mixtures were prepared by mass using small vessels of about 10 cm$^3$. The entire process was carefully handled to prevent evaporation. The uncertainty of the final mole fraction is ± 0.0001. Molar quantities were obtained using the relative atomic mass table of 2015 issued by I.U.P.A.C [40].

Measurements were carried out at atmospheric pressure and at (293.15, 298.15, 303.15) K. An Anton Paar DMA 602 vibrating-tube densimeter was used for the determination of $\rho$ values. The apparatus was calibrated at each temperature. Details on the calibration method and on the measurement protocol can be found elsewhere [41,42]. The resolution in $\rho$ is $|\Delta\rho/\rho| =$ 6 10$^{-6}$, apart from the errors due to calibration and the density of the reference liquids. The accuracy of the density measurements is estimated to be ± 5·10$^{-2}$ kg.m$^{-3}$. The temperature stability of the densimeter, achieved by means of a LAUDA RE 304 heating bath, was ± 0.01 K. For $V_m^E$, the estimated accuracy is less than ± (0.01 $|V_{m,min}^E|$ + 0.005 cm$^3$·mol$^{-1}$), where $V_{m,min}^E$ stands for the minimum experimental value of $V_m^E$ with respect to the mole fraction.

Values of kinematic viscosities were obtained from an Ubbelohde viscosimeter with a Schoot-Geräte automatic measuring unit model AVS-350. The temperature was kept constant within ± 0.01 K by means of a controller bath CT52, also from Schott. Apparatus calibration has been explained in detail previously [43] and was also conducted at each temperature. The uncertainties for $\nu$ and $\eta$ values are estimated to be ± 1%. For the deviations of viscosity from linear dependence on molar fraction (see next Section), the corresponding uncertainty is ± 2%. These estimations are confirmed by the comparison of our viscosity results with literature values for the 1-propanol + CH$_3$OCOOCH$_3$ system at different temperatures [44].

3     EXPERIMENTAL RESULTS AND CORRELATIONS

Table S1 (supplementary material) lists values of $\rho$ and $\eta$ vs. $x_1$, the mole fraction of methanol. We have also computed $V_m^E$ and deviations of $\eta$ from the linear dependence on mole fraction (Table S1, supplementary material) using the equation:

$$\Delta \eta = \eta - (x_1 \eta_1 + x_2 \eta_2) \qquad (1)$$

The $V_m^E$ and $\Delta \eta$ data (Figures 1 and 2) have been fitted by unweighted least-squares polynomial regression to the equation:

$$\Delta F = x_1(1-x_1) \sum_{i=0}^{k-1} A_i (2x_1 - 1)^i \qquad (2)$$

being $\Delta F = V_m^E, \Delta \eta$. The number, $k$, of coefficients $A_i$ used in equation (2) for each system was determined by applying an F-test [45] at the 99.5 % confidence level. Table 3 lists the parameters $A_i$ obtained in the regression, together with the standard deviations $\sigma$, defined by:

$$\sigma(\Delta F) = \left[ \frac{1}{N-k} \sum \left( \Delta F_{cal} - \Delta F_{exp} \right)^2 \right]^{1/2} \qquad (3)$$

where $N$ is the number of direct experimental points.

*3.1 Comparison with data available in the literature*

The present $V_m^E$ results at 298.15 K are in good agreement with those previously provided by us [11] (Figure 1). However, $V_m^E$ data from reference [28] for the system including 2,5,8-trioxanonane are slightly lower than our values (Figure 1). Thus, at equimolar composition and 298.15 K, $V_m^E$/cm$^3$.mol$^{-1}$ = $-0.5978$ (this work) and $-0.624$ [28]. In contrast, our $V_m^E$ data for the system involving 2,5,8,11,14-pentaoxapentadecane at 293.15 K and 303.15 K agree well with the measurements available in the literature. At $x_1$ = 0.5, we have $V_m^E$/cm$^3$.mol$^{-1}$ = $-0.7778$ (293.15 K) and $-0.7891$ (303.15 K). Literature values, in the same units, are: $-0.781$ [29] and $-0.789$ [30] at 293.15 K and $-0.792$ [29] and $-0.779$ [30] at 303.15 K. It is important to underline that the magnitude $A_p = (\frac{\partial V^E}{\partial T})_p$ for the system with the pentaether from reference [30] is positive at 298.15 K ($A_p(x_1 = 0.5)$/cm$^3$.mol$^{-1}$.K$^{-1}$ = 1.05 10$^{-3}$). Our measurements indicate that

$A_p$ is negative ($A_p(x_1 = 0.5)/\text{cm}^3.\text{mol}^{-1}.\text{K}^{-1} = -1.1 \, 10^{-3}$) which is confirmed by results from reference [29] ($-1.1 \, 10^{-3} \, \text{cm}^3.\text{mol}^{-1}.\text{K}^{-1}$ at the same conditions). The accurate determination of $A_p$ values (see below) is crucial since the sign of this magnitude is closely related to interactional and structural effects (see below). The $\Delta\eta$ values for the methanol + 2,5,8,11,14-pentaoxapentadecane system at 293.15 K reported in reference [29] largely differs from the results provided here. Thus, $\Delta\eta$ ($x_1 = 0.5$) = 0.181 mPa.s [29] is much lower than our result (0.372 mPa.s). This may be due to, at the mentioned temperature, the viscosity of pure methanol given in reference [29], (0.985 mPa.s) is very different to the experimental values encountered in the literature (Table 2). In contrast, at 303.15 K, the $\Delta\eta$ ($x_1 = 0.5$) results for this system agree well (0.334 [29] and 0.332 (this work) mPa.s).

*3.2 Viscosity correlations*

Expressions for all the mentioned equations are written in Supplementary Material. The equations used for the correlation of dynamic viscosities have an adjustable parameter termed as follows: $G_{12}$ (Grunberg-Nissan) [31], $\eta_{12}$ (Hind) [32], $F_{12}$ (Frenkel) [33], and $W_{12}$ (Katti-Chaudhri) [34]. In the case of kinematic viscosities, the applied equations have two adjustable parameters: $Z_{12}, Z_{21}$ (McAllister equation) [35] and $\alpha_{12}, \alpha'_{12}$ (Heric equation) [36]. The McAllister equation results from the combination of the Eyring's theory of absolute reaction rates and a three-body interaction model [35,46].

Results obtained using the mentioned equations are compared by means of the relative standard deviations $\sigma_r$. They are calculated according to the expression:

$$\sigma_r(F) = \left[ \frac{1}{N-k} \sum \left( \frac{F_{cal} - F_{exp}}{F_{exp}} \right)^2 \right]^{1/2} \quad (4)$$

being $F = \eta$ or $\nu$. Values of the fitted parameters for the applied equations together with the corresponding $\sigma_r$ values are compiled in Table 4. From inspection of this Table, we can conclude: (i) $\sigma_r$ values increase with the ether size, i.e., when structural effects become more important (see below); (ii) better results are obtained from the Hind equation; (iii) the use of equations with two adjustable parameters does not lead to improved results.

## 4. ERAS MODEL

A brief summary of the model is now provided. (i) The excess functions are determined from the equation:

$$F_{m}^{E} = F_{m,chem}^{E} + F_{m,phys}^{E} \qquad (5)$$

In this work, we consider $F = H$ (enthalpy), or $V$ (volume). $F_{m,chem}^{E}$ is the chemical contribution and arises from hydrogen-bonding effects. $F_{m,phys}^{E}$ represents the physical contribution, consequence of the existence of non-polar Van der Waals' interactions and free volume effects. The needed expressions for the $H_{m}^{E}$ and $V_{m}^{E}$ calculations can be found elsewhere [47,48]. (ii) It is assumed that self-associated species only form linear chains. This association process is described by a chemical equilibrium constant ($K_A$), which does not depend on the chain length of the associated compound (methanol, in the present work), according to the equation:

$$A_m + A \leftrightarrow A_{m+1} \qquad (6)$$

where $m$ varies from 1 to $\infty$. The cross-association between a self-associated compound $A_m$ and a non self-associated component $B$ (here, linear polyethers) is represented by

$$A_m + B \xleftrightarrow{K_{AB}} A_m B \qquad (7)$$

The cross-association constants ($K_{AB}$) are also assumed to be independent of the chain length. Equations (6) and (7) are characterized by $\Delta h_i^*$, the enthalpy of the reaction that corresponds to the hydrogen-bonding energy, and by the volume change ($\Delta v_i^*$) related to the formation of the linear chains. (iii) The $F_{m,phys}^{E}$ term is derived from the Flory's equation of state [26], which is assumed to be valid not only for pure compounds but also for the mixture [49,50]:

$$\frac{\overline{P}_i \overline{V}_i}{\overline{T}_i} = \frac{\overline{V}_i^{1/3}}{\overline{V}_i^{1/3} - 1} - \frac{1}{\overline{V}_i \overline{T}_i} \qquad (8)$$

where i = A,B or M (mixture). In equation (8), $\bar{V}_i = V_i/V_i^*$; $\bar{P}_i = P/P_i^*$; $\bar{T}_i = T/T_i^*$ are the reduced properties for volume, pressure and temperature, respectively. The reduction parameters of the pure compounds ($V_i^*, P_i^*, T_i^*$) are obtained from P-V-T data ($\rho$, $\alpha_p$, isobaric thermal expansion coefficient, and isothermal compressibility, $\kappa_T$), and association parameters [49,50]. The reduction parameters of the system, $P_M^*$ and $T_M^*$, are calculated following well known mixing rules [49,50]. The total relative molecular volumes and surfaces of the compounds were calculated additively on basis the group volume and surface values recommended by Bondi [51].

### 4.1 Adjustment of ERAS parameters

Values of $V_i$, $V_i^*$ and $P_i^*$ of linear polyethers at 298.15 K, needed for calculations, have been taken from a previous work [14]. Similarly, for methanol, the values used are the same as those given in [52]. On the other hand, for this alcohol, $K_A$ (= 986) $\Delta h_A^*$ (= − 25.1 kJ.mol$^{-1}$) and $\Delta v_A^*$ (= − 5.6 cm$^3$.mol$^{-1}$) are values commonly used in many different studies [48,52,53]. The binary parameters $K_{AB}, \Delta h_{AB}^*, \Delta v_{AB}^*$ and $X_{AB}$ are then fitted against $H_m^E$ [15,54,55] and $V_m^E$ [11, this work] data available in the literature for methanol + linear polyether systems. These parameters are collected in Table 5.

### 4.2 Results

ERAS results are graphically shown along Figures 3-5. We must underline that the model provides acceptable $H_m^E$ results. In contrast, the shape of the $V_m^E(x_1)$ curves is not properly described.

## 5. DISCUSSION

Below, we are referring to the values of the thermophysical properties at equimolar composition and 298.15 K. On the other hand, $n$ indicates the number of CH$_2$CH$_2$O groups in the linear polyether

### 5.1 Calorimetric data

The $H_m^E$/J.mol$^{-1}$ values of methanol + CH$_3$-O-(CH$_2$CH$_2$O)$_n$-CH$_3$ mixtures are positive and rather low. They change in the sequence: 338 ($n$ = 1) [15] < 440 ($n$ = 2) [54] < 581 ($n$ = 4) [55]. Therefore, the contribution to $H_m^E$ from the breaking of interaction between like molecules is dominant over the negative contribution due to the creation of alkanol-ether interactions along the mixing process. In addition, $H_m^E$ increases in line with $n$. This may be explained taking into account that the contribution to $H_m^E$ from the disruption of ether-ether interactions also increases in line with $n$. In fact, for $n$-C$_7$ mixtures, we have $H_{\text{m}}^{\text{E}}$/J.mol$^{-1}$ = 1285 ($n$ = 1) [56] < 1621 ($n$ = 2) [57] < 1754 ($n$ =3) [58] < 1891 ($n$ = 4) [57]. It is to be noted that these values are much higher than those given above for the corresponding methanol systems, which underlines the importance of the alkanol-ether interactions created upon mixing. In the same sense, we must underline that methanol + heptane system shows a miscibility gap at 298.15, with an upper critical solution temperature at 324.1 K [59].

### 5.2 Volumetric data

The $V_{\text{m}}^{\text{E}}$/cm$^3$.mol$^{-1}$ values of the investigated mixtures are rather large and negative. The variation with $n$ is as follows: $-0.4991$ ($n$ = 1) [11] > $-0.5978$ ($n$ =2) > $-0.7155$ ($n$ = 3) > $-0.7834$ ($n$ = 4) (this work). Interestingly, the signs of both excess functions $H_{\text{m}}^{\text{E}}$ and $V_{\text{m}}^{\text{E}}$ are opposite. This is the typical behaviour of mixtures characterized by structural effects [60]. It is remarkable that, while $H_{\text{m}}^{\text{E}}$ increases in line with $n$, $V_{\text{m}}^{\text{E}}$ decreases. That is, structural effects become more important with the number of CH$_2$CH$_2$O groups involved in the ether. The mentioned structural effects may be of free volume type, since the difference between $\alpha_p$ values of the two mixture components also increases in line with $n$ [61]. Thus, ($\alpha_p$ (methanol) $-\alpha_p$ (ether))/10$^{-3}$.K$^{-1}$: $-0.072$ ($n$ =1) < 0.136 ($n$ = 2) < 0.231 ($n$ = 3) < 0.275 ($n$ = 4) [62,63]. In addition, the $V_{\text{m}}^{\text{E}}$ curves become skewed towards higher mole fractions of methanol when the oxaalkane size increases (Figure 1), a normal behaviour of systems where structural effects exist. It is remarkable that, in the present case, structural effects have a great impact on $H_{\text{m}}^{\text{E}}$. It is well-known that this magnitude is determined by both interactional and structural effects [64,65]. Interactional effects are more adequately considered by means of the excess internal energy at constant volume, $U_{\text{Vm}}^{\text{E}}$. If terms of higher order in $V_{\text{m}}^{\text{E}}$ are neglected, $U_{\text{Vm}}^{\text{E}}$ can be written as [64,65]:

$$U_{\text{Vm}}^{\text{E}} = H_{\text{m}}^{\text{E}} - \frac{\alpha_p}{\kappa_T} T V_{\text{m}}^{\text{E}} \qquad (9)$$

In this expression, $\frac{\alpha_p}{\kappa_T} T V_m^{\text{E}}$ is the equation of state (eos) contribution to $H_{\text{m}}^{\text{E}}$, and $\alpha_p$ and $\kappa_T$ are, respectively, the isobaric thermal expansion coefficient and the isothermal compressibility of the mixture Here, we have determined $\alpha_p$ and $\kappa_T$ assuming ideal behavior ($F = \varphi_1 F_1 + \varphi_2 F_2$; $F_i$ is the property of the pure compound $i$ for these magnitudes [14,52] and $\varphi_i$ is the volume fraction). In fact, calculations show that $\alpha_p$ values for the present systems are close to the ideal ones. Thus, for mixtures containing methanol, $U_{Vm}^E$/J.mol$^{-1}$ = 501 (2,5-dioxahexane); 655 (2,5,8-trioxanonane); 910 (2,5,8,11,14-pentaoxapentadecane, values which largely differ from the $H_m^E$ results.

The magnitude $A_p = (\frac{\partial V^E}{\partial T})_p$ is useful to gain insights into the interactional and structural features of the systems under consideration. Its sign is the result of the variation in the balance of association/solvation and structural effects with temperature [66]. For systems formed by a short 1-alkanol and a long $n$-alkane, we have $A_p > 0$ over the whole concentration range. Association effects are then dominant. If the solution contains a long 1-alkanol and a short $n$-alkane, $A_p$ is negative at the concentration region where interstitial accommodation becomes important [66-68]. For linear polyether + alkane mixtures, $A_p$ is usually positive (i.e., interactional effects, related to dipolar interactions, are determinant) and decreases with the oxaalkane size, being negative for the 2,5,8,11,14-pentaoxapentadecane + heptane or + cyclohexane systems [69]. In such a case, structural effects are dominant. For the present systems, and at the compositions where the $V_m^E$ curves have their minimum value, we have determined $A_p$ from linear regressions (regression coefficients = 1) for the $\rho(T)$ data to obtain the corresponding $(\frac{\partial \rho}{\partial T})_p$ values at the desired compositions. The results are $A_p$ /cm$^3$.mol$^{-1}$.K$^{-1}$ = $-7.3 \cdot 10^{-4}$ ($n = 2$: $x_1 = 0.5993$); $-1.3 \cdot 10^{-3}$ ($n = 3$; $x_1 = 0.5962$); $-2.1 \cdot 10^{-3}$ ($n = 4$; $x_1 = 0.6966$). This confirms the existence of structural effects in the present solutions. The negative $A_p$ (303.15 K) value of the 1-propanol + PEG-250 mixture ($-0.001$ cm$^3$.mol$^{-1}$.K$^{-1}$ [70]) is in agreement with our statement. In systems with 2,5-dioxahexane, the change from a positive $A_p$ value for the ethanol mixture ($A_p = 2.4 \cdot 10^{-3}$ cm$^3$.mol$^{-1}$.K$^{-1}$) to a negative value for the 1-octanol solution ($-3.2 \cdot 10^{-3}$ cm$^3$.mol$^{-1}$.K$^{-1}$ [71]) reveals that association effects are dominant in the former mixture while structural effects are determinant in the latter. It is also interesting to compare $A_p$ results for methanol + 2,5,8-trioxanonane, or + 3,6,9-trioxaundecane (2 $10^{-3}$ cm$^3$.mol$^{-1}$.K$^{-1}$ [14]) systems. Such difference in the sign newly reveals that association effects are more important in the 3,6,9-trioxaundecane solution, since the O atoms are more sterically hindered in this polyether and alkanol-ether are more difficult to be created.

*5.3 Viscometric data*

The $\Delta \eta$ values listed in S1 of supplementary material are positive and increase with $n$: 0.068 ($n = 2$) < 0.195 ($n = 2$) < 0.349 ($n = 4$), all values in mPa.s. These positive $\Delta \eta$ results reveal that there is a loss of fluidization of the system when the mixing is produced, which is usually ascribed to the existence of strong interactions between unlike molecules [72]. For example, $\Delta \eta$ /mPa.s = 0.13 (methanol + 1-propylamine) [73], 0.20 (methanol + 1-butylamine) [74], 0.18 (CHCl$_3$ + *N,N,N*-triethylamine (TEA)) [75]. For the cited solutions, their $H_m^E$ values largely differ from the results given above for methanol + polyether systems. Thus, $H_m^E$/J.mol$^{-1}$ = $-$3794 (methanol + 1-propylamine) [49], $-$3767 (methanol + 1-butylamine) [49], $-$4072 (CHCl$_3$ + TEA) [76] Negative $\Delta \eta$ values are commonly interpreted in terms of a higher fluidization of the system caused by the disruption of interactions between like molecules [77], which overcompensates the increase of $\Delta \eta$ related to the creation of interactions between unlike molecules upon mixing [43]. This trend is clearly observed for 1-alkanol + alkane, or + amine mixtures [43,77-79]. However, viscosity is a very sensitive magnitude to size effects and our results should be attributed to the existence of strong structural effects. In fact, mixtures characterized by strong interactions between unlike molecules and positive $\Delta \eta$ results may show $\eta(x_1)$ curves with a maximum [72], a behaviour not observed for the treated solutions. Finally, it should be mentioned that using the value $\Delta \eta = -0.036$ mPa.s for the methanol + 2,5-dioxahexane mixture [80], we find a good correlation between $\Delta \eta$ and the difference in volume between the mixture components, $V_{m2} - V_{m1}$, for methanol + CH$_3$-O-(CH$_2$CH$_2$O)$_n$-CH$_3$ systems. Thus, $\Delta \eta$ = ($-$0.255 mPa.s) + (0.0033 mPa.s/cm$^3$.mol$^{-1}$)($V_{m2} - V_{m1}$), with $r$ = 0.994 and the standard deviation equal to 0.015 mPa.s.

It is known that the application of the Grunberg-Nissan equation to the correlation of viscosity data of systems characterized by positive deviations from the Raoult's law leads to the determination of small and negative $G_{12}$ values. In contrast, positive $G_{12}$ values are obtained for systems which show negative deviations from the Raoult's law [72,75]. In our case, $G_{12}$ = 0.515 ($n$ = 2), 1.280 ($n$ =3); 1.972 ($n$ = 4) (Table 4). 1-Alkanol + polyether mixtures deviate positively from the Raoul's law as it is indicated by the corresponding values of the molar excess Gibbs energies, $G_m^E$. Thus, for mixtures containing 2,5,8-trioxanonane ($T$ = 308.15 K), $G_m^E$/J.mol$^{-1}$ = 283 (methanol), 313 (1-propanol) [81]. In consequence, the change of the expected sign of $G_{12}$ should be ascribed to the existence of strong structural effects. Note that, newly, there is good linear dependence of $G_{12}$ with $V_{m2} - V_{m1}$ since $G_{12}$ = $-$ 1.377 + (0.019 cm$^{-3}$.mol)($V_{m2} - V_{m1}$), with $r$ = 0.999. Similar results have been obtained for other systems with compounds which differ largely in size. For example, $G_{12}$ = 1.95 for the 1-decanol + 1-propylamine mixture at 303.15 K [82].

As usually, $\Delta\eta$ decreases with the increasing of temperature (Table S1, supplementary material). The $\rho$ and $\eta$ data allow compute the Gibbs energy of activation of viscous flow, $\Delta G^*$, in the framework of the Eyring's theory [46,83,84]. The enthalpy, $\Delta H^*$, and entropy, $\Delta S^*$, of activation of viscous flow can be calculated from the mentioned data at different temperatures using the expression [85,86]:

$$\ln \frac{\eta V_m}{hN_A} = \frac{\Delta H^*}{RT} - \frac{\Delta S^*}{R} \qquad (10)$$

The plots of $\ln(\eta V/hN_A)$ against $1/T$ give a straight line for each mixture. The magnitudes $\Delta H^*$, and $\Delta S^*$ can be then estimated from its slope and intercept. Results on $\Delta H^*$, $\Delta S^*$ and $\Delta G^*$ are listed in Table S2 (supplementary material). Data reveal that $\Delta S^*$, at a given composition, is rather small during the activated viscous flow process and that $\Delta G^*$ is essentially determined by enthalpic effects. On the other hand, $\Delta H^*$ increases with $n$, i.e., when the solvent is more structured since dipolar interactions are more relevant in the pentaether system than in the triether solution. This is supported by the relative variation of $H_m^E$ values and of UCSTs of $CH_3$-O-$(CH_2CH_2O)_n$-$CH_3$ + $n$-alkane systems (see above). In addition, it is remarkable, that $\Delta S^*$ of pure oxaalkane become less negative when its chain length increases (Table S2, supplementary material). We have also determined $\Delta(\Delta G^*)$ ($=\Delta G^* - x_1\Delta G_1^* - x_2\Delta G_2^*$) at 298.15 K (Table S2, supplementary material and Figure 6). The $\Delta(\Delta G^*)$ /J·mol$^{-1}$ variation is as follows: 759 ($n = 2$) < 1465 ($n = 3$) < 2086 ($n = 4$), and also changes linearly with $V_{m2} - V_{m1}$. Large positive values of $\Delta(\Delta G^*)$ have been also encountered for mixtures with compounds very different in size, as for the already mentioned system 1-decanol + 1-propylamine at 303.15 K) (= 1478 J.mol$^{-1}$) [82].

*5.4   ERAS results*

We have presented, previously, a detailed study on 1-alkanol + linear polyether systems in terms of the Flory model [14]. One of the conclusions from this research was that orientational effects play a crucial role in systems with methanol or ethanol, and that the random mixing approximation is attained, in large extent, for the remainder mixtures. For this reason, we explore here the ability of ERAS to describe simultaneously $H_m^E$ and $V_m^E$ of methanol-containing systems. The ERAS parameters listed in Table 5 deserve some comments. (i) The $\Delta h_{AB}^*$ values are practically equal to those determined earlier from calorimetric data. Thus, we have obtained the following results for the enthalpy of hydrogen bond, $\Delta H_{OH-O}$, for alkanol-ether interactions in 1-alkanol + linear polyether systems [14]: −26.4 (2,5-dioxahexane); −26.3 (3,6-dioxaoctane);

− 28.6 (2,5,8-trioxanonane) and − 34.2 (2,5,8,11,14-pentaoxapentadecane) (values in kJ.mol$^{-1}$). (ii) The large and negative $\Delta v_{AB}^*$ values may be considered to be due to large structural effects, more important in the pentaether solution. (iii). We note that $K_{AB}$ is lower and $X_{AB}$ is higher for the 2,5,8,11,14-pentaoxapentadecane system compared with the values of the 2,5-dioxahexane mixture. Therefore, it seems that physical interactions are more important in the solution with the pentaether. The same trend is observed when comparing $K_{AB}$ and $X_{AB}$ results listed in Table 5 with those for methanol + linear monoether mixtures. Thus, for the methanol + di-*n*-propylether system, $K_{AB}$ = 54 and $X_{AB}$ = 3.6 J.cm$^{-3}$ [7]. It is clear that cross-association effects are more relevant in mixtures with linear monoethers. (iv) Similarly, solvation effects are weaker in the methanol + 3,6-dioxaoctane mixture ($K_{AB}$ = 14.8) compared to those in the corresponding system with 2,5-dioxahexane ($K_{AB}$ = 18) (Table 5). In contrast, physical interactions are more relevant in the latter mixture since it is characterized by a higher $X_{AB}$ parameter (Table 5). As a rule, acetal-acetal interactions are weaker than those between similar ether molecules [87]. For example, for mixtures with heptane, $H_m^E$/J.mol$^{-1}$ = 1285 (2,5-dioxahexane) [56] > 889 (3,6-dioxaoctane) [88]. Finally, we underline that results shown in Figures 3-5 reveal that the simultaneous representation of $H_m^E$ and $V_m^E$ is a hard task for the model, which can be explained in terms existence of strong dipolar interactions and large structural effects.

The relevance of physical interactions can be demonstrated examining the position of these systems within the $G_m^E$ vs. $H_m^E$ diagram [89-91]. The main features of this diagram can be found elsewhere [92]. Taking into account the values of $G_m^E$ and $H_m^E$ of the methanol + 2,5,8-trioxanonane system given above, we note that this solution is placed between the lines $G_m^E = H_m^E/2$ and $G_m^E = H_m^E$, the region where are situated mixtures characterized by dipolar interactions (alkanone, or alkanoate, or dialkyl organic carbonate + alkane). For the 1-propanol + 2,5,8-trioxanonane system, $H_m^E$ = 1214 J.mol$^{-1}$ [54] and it is situated close to the region where solutions characterized by dispersive interactions are encountered (between the lines $G_m^E = H_m^E/3$ and $G_m^E = H_m^E/2$.). Accordingly, Flory results are much better for this system than for the corresponding methanol mixture [14].

## 6. CONCLUSIONS

Density and viscosity data have been determined for the systems methanol + $CH_3$-O-$(CH_2CH_2O)_n$-$CH_3$ ($n$ =2,3,4) at (293.15-303.15) K . Derived $V_m^E$ results indicate the existence of large structural effects, of the free volume type, which increase in line with $n$. The relevance of structural effects is confirmed by the viscosity measurements or by the results obtained from the correlation of the data using the Grunberg-Nissan equation. The application of the ERAS model to methanol + linear polyether mixtures newly supports the existence of strong physical interactions and structural effects in such systems.

## 7. ACKNOWLEDGEMENTS

The authors gratefully acknowledge the financial support received from the Consejería de Educación y Cultura de Castilla y León, under Project VA100G19 (Apoyo a GIR, BDNS: 425389).

TABLE 1

Source and purity of the pure compounds used in this work

| Compound | Source | Purity[a] | Methods of purification |
|---|---|---|---|
| Methanol | Sigma-Aldrich | 0.998 | Dried with molecular sieves Type 3A |
| 2,5,8-trioxanonane | Fluka | ≥ 0.99 | Dried with molecular sieves Type 3A |
| 2,5,8,11-tetraoxadodecane | Fluka | ≥ 0.98 | Dried with molecular sieves Type 3A |
| 2,5,8,11,14-pentaoxapentadecane | Fluka | ≥ 0.98 | Dried with molecular sieves Type 3A |

[a]in mole fraction (by gas chromatography. Provided by the supplier)

TABLE 2

Thermophysical properties of pure compounds at temperature $T$ and 0.1 MPa: $\rho$, density and $\eta$, dynamic viscosity[a]

| Compound | T/K | $\rho$ /g.cm$^{-3}$ | | $\eta$ /mPa.s | |
|---|---|---|---|---|---|
| | | Exp. | Lit. | Exp. | Lit. |
| Methanol | 293.15 | 0.791576 | 0.79190[b] | 0.590 | 0.585[b] |
| | | | 0.79109[c] | | 0.593[c] |
| | | | 0.79154[d] | | 0.582[d] |
| | 298.15 | 0.786872 | 0.78720[b] | 0.551 | 0.545[b] |
| | | | 0.78648[c] | | 0.551[c] |
| | | | 0.78683[d] | | 0.550[d] |
| | 303.15 | 0.782140 | 0.78248[b] | 0.505 | 0.508[b] |
| | | | 0.78122[c] | | 0.509[c] |
| | | | 0.78212[d] | | 0.512[d] |
| 2,5,8-trioxanonane | 293.15 | 0.944138 | 0.9434[e] | 1.079 | 1.09[f] |
| | | | 0.9452[f,g] | | 1.081[g] |
| | 298.15 | 0.939171 | 0.9385[e] | 0.997 | 0.998[f] |
| | | | 0.9398[f] | | 1.011[g] |
| | | | 0.9402[g] | | |
| | 303.15 | 0.934206 | 0.9335[e] | 0.916 | 0.914[f] |
| | | | 0.9350[f] | | 0.942[g] |
| | | | 0.9352[g] | | |
| 2,5,8,11-tetraoxadodecane | 293.15 | 0.986165 | 0.98416[h] | 2.188 | 2.19[i] |
| | | | 0.9860[g] | | 2.162[g] |
| | | | 0.98569[j] | | 2.158[j] |
| | 298.15 | 0.981408 | 0.97961[h] | 1.974 | 1.96[i] |
| | | | 0.9815[g] | | 1.956[g] |
| | | | 0.98123[k] | | |
| | 303.15 | 0.976657 | 0.97485[h] | 1.777 | 1.78[i] |
| | | | 0.9767[g] | | 1.777[g] |
| | | | 0.97672[j] | | 1.761[j] |
| 2,5,8,11,14-pentaoxpanetadecane | 293.15 | 1.011350 | 1.012060[l] | 3.817 | 3.84[i] |
| | | | 1.0123[g] | | 3.846[g] |
| | | | 1.0116[j] | | 3.795[l] |
| | 298.15 | 1.006723 | 1.00726[k] | 3.378 | 3.40[i] |
| | | | 1.0076[g] | | 3.335[g] |

TABLE 2 (continued)

| | | | | | |
|---|---|---|---|---|---|
| | | | 1.00742[l] | | 3.348[l] |
| | | | 1.0063[m] | | 3.294[m] |
| | | | 1.0070[n] | | 3.334[n] |
| | 303.15 | 1.002115 | 1.0020[n] | 2.985 | 2.955[n] |
| | | | 1.0031[g] | | 2.967[g] |
| | | | 1.0029[j] | | 2.951[j] |
| | | | 1.00279[l] | | 2.983[l] |
| | | | | | 2.99[i] |

[a]uncertainties, $u$, are: $u(T) = \pm\ 0.01$ K; $u(p) = 1$ kPa; $u(\rho) = \pm\ 5\cdot10^{-5}$ g·cm$^{-3}$; $u(\eta) = \pm\ 0.01\cdot\eta$ mPa·s; [b][93]; [c][86]; [d][94]; [e][95]; [f][96]; [g][97]; [h][98]; [i][99]; [j][100]; [k][101]; [l][102]; [m][103]; [n][104]

TABLE 3

Coefficients $A_i$ and standard deviations, $\sigma(\Delta F)$ (eq. 3) for representation of the $\Delta F$ property at temperature $T$ for methanol(1) + linear polyether(2) systems at temperature, $T$ and 0.1 MPa by eq. 2.

| Property: $\Delta F$ | $T$/K | $A_0$ | $A_1$ | $A_2$ | $A_3$ | $\sigma(\Delta F)$ |
|---|---|---|---|---|---|---|
| Methanol(1) + 2,5,8-trioxanonane(2) | | | | | | |
| $V_m^E$ /cm$^3$.mol$^{-1}$ | 293.15 | $-2.3779$ | $-0.9656$ | $-0.6524$ | | 0.006 |
| $\Delta\eta$ /mPa.s | | 0.294 | 0.099 | 0.029 | | 0.001 |
| $V_m^E$ /cm$^3$.mol$^{-1}$ | 298.15 | $-2.3912$ | $-0.9891$ | $-0.6513$ | | 0.006 |
| $\Delta\eta$ /mPa.s | | 0.273 | 0.092 | 0.039 | | 0.001 |
| $\Delta(\Delta G^*)$ /J.mol$^{-1}$ | | 3037 | 1351 | 590 | | 6 |
| $V_m^E$ /cm$^3$.mol$^{-1}$ | 303.15 | $-2.3986$ | $-1.0164$ | $-0.6630$ | | 0.006 |
| $\Delta\eta$ /mPa.s | | 0.249 | 0.094 | 0.040 | | 0.001 |
| Methanol(1) + 2,5,8,11-tetroxadodecane(2) | | | | | | |
| $V_m^E$ /cm$^3$.mol$^{-1}$ | 293.15 | $-2.8319$ | $-1.2084$ | $-0.8808$ | $-0.8077$ | 0.005 |
| $\Delta\eta$ /mPa.s | | 0.839 | 0.158 | $-0.180$ | | 0.002 |
| $V_m^E$ /cm$^3$.mol$^{-1}$ | 298.15 | $-2.8620$ | $-1.2130$ | $-0.8743$ | $-0.8594$ | 0.005 |
| $\Delta\eta$ /mPa.s | | 0.781 | 0.172 | $-0.146$ | | 0.002 |
| $\Delta(\Delta G^*)$ /J.mol$^{-1}$ | | 5862 | 3235 | 1552 | | 13 |
| $V_m^E$ /cm$^3$.mol$^{-1}$ | 303.15 | $-2.8947$ | $-1.2076$ | $-0.8961$ | $-0.9058$ | 0.0053 |
| $\Delta\eta$ /mPa.s | | 0.733 | 0.186 | $-0.117$ | | 0.002 |
| Methanol(1) + 2,5,8,11,14-pentaoxapentadecane(2) | | | | | | |
| $V_m^E$ /cm$^3$.mol$^{-1}$ | 293.15 | $-3.1115$ | $-1.5680$ | $-1.8493$ | $-0.8802$ | 0.008 |
| $\Delta\eta$ /mPa.s | | 1.487 | $-0.117$ | $-0.306$ | | 0.004 |
| $V_m^E$ /cm$^3$.mol$^{-1}$ | 298.15 | $-3.1336$ | $-1.6083$ | $-1.9202$ | $-0.8543$ | 0.008 |
| $\Delta\eta$ /mPa.s | | 1.399 | $-0.063$ | $-0.203$ | | 0.005 |
| $\Delta(\Delta G^*)$ /J.mol$^{-1}$ | | 8346 | 4341 | 3138 | 2023 | 7 |
| $V_m^E$ /cm$^3$.mol$^{-1}$ | 303.15 | $-3.1564$ | $-1.6454$ | $-1.9713$ | $-0.8831$ | 0.008 |
| $\Delta\eta$ /mPa.s | | 1.326 | 0.024 | $-0.161$ | | 0.004 |

TABLE 4

Fitted parameters and relative standard deviations, $\sigma_r$ for the semi-empirical equations used for the correlation of viscosity data for methanol(1) + linear polyether(2) mixtures at temperature $T$.

| Equation | Parameter | $\sigma_r$ [a] | Parameter | $\sigma_r$ [a] | Parameter | $\sigma_r$ [a] |
|---|---|---|---|---|---|---|
| | $T = 293.15$ K | | $T = 298.15$ K | | $T = 303.15$ K | |
| | Methanol(1) + 2,5,8-trioxanonane(2) | | | | | |
| Grunberg-Nissan | 0.518 | 0.015 | 0.515 | 0.014 | 0.514 | 0.013 |
| Hind | 0.985 | 0.007 | 0.915 | 0.007 | 0.840 | 0.007 |
| Frenkel | 1.034 | 0.015 | 0.959 | 0.014 | 0.880 | 0.013 |
| Katti-Chaudhri[b] | 1.290 | 0.037 | 1.285 | 0.037 | 1.285 | 0.038 |
| McAllister | $Z_{12} = 1.231$ | 0.008 | $Z_{12} = 1.149$ | 0.009 | $Z_{12} = 1.066$ | 0.009 |
| | $Z_{21} = 1.049$ | | $Z_{21} = 0.977$ | | $Z_{21} = 0.898$ | |
| Heric | $\alpha_{12} = 0.818$ | 0.008 | $\alpha_{12} = 0.810$ | 0.008 | $\alpha_{12} = 0.824$ | 0.009 |
| | $\alpha'_{12} = 0.093$ | | $\alpha'_{12} = 0.087$ | | $\alpha'_{12} = 0.117$ | |
| | Methanol(1) + 2,5,8,11-tetraoxadodecane(2) | | | | | |
| Grunberg-Nissan | 1.302 | 0.072 | 1.280 | 0.066 | 1.282 | 0.061 |
| Hind | 1.796 | 0.012 | 1.644 | 0.012 | 1.502 | 0.013 |
| Frenkel | 2.178 | 0.072 | 1.978 | 0.066 | 1.800 | 0.061 |
| Katti-Chaudhri[b] | 2.509 | 0.090 | 2.485 | 0.089 | 2.484 | 0.091 |
| McAllister | $Z_{12} = 2.637$ | 0.020 | $Z_{12} = 2.426$ | 0.020 | $Z_{12} = 2.242$ | 0.021 |
| | $Z_{21} = 1.763$ | | $Z_{21} = 1.604$ | | $Z_{21} = 1.449$ | |
| Heric | $\alpha_{12} = 2.758$ | 0.020 | $\alpha_{12} = 2.729$ | 0.020 | $\alpha_{12} = 2.754$ | 0.021 |
| | $\alpha'_{12} = 1.598$ | | $\alpha'_{12} = 1.595$ | | $\alpha'_{12} = 1.644$ | |
| | Methanol(1) + 2,5,8,11,14-pentaoxapentadecane(2) | | | | | |
| Grunberg-Nissan | 2.007 | 0.176 | 1.972 | 0.156 | 1.962 | 0.144 |

TABLE 5 (continued)

| | | | | | | |
|---|---|---|---|---|---|---|
| Hind | 2.920 | 0.014 | 2.647 | 0.009 | 2.395 | 0.006 |
| Frenkel | 4.090 | 0.176 | 3.657 | 0.156 | 3.277 | 0.144 |
| Katti-Chaudhri[b] | 3.675 | 0.150 | 3.637 | 0.149 | 3.616 | 0.151 |
| McAllister | $Z_{12} = 5.604$ | 0.042 | $Z_{12} = 5.046$ | 0.043 | $Z_{12} = 4.622$ | 0.044 |
| | $Z_{21} = 2.623$ | | $Z_{21} = 2.358$ | | $Z_{21} = 2.093$ | |
| Heric | $\alpha_{12} = 4.769$ | 0.042 | $\alpha_{12} = 4.695$ | 0.043 | $\alpha_{12} = 4.712$ | 0.044 |
| | $\alpha'_{12} = 3.309$ | | $\alpha'_{12} = 3.258$ | | $\alpha'_{12} = 3.313$ | |

[a]eq. (4); [b]values of $W_{12}/RT$ ($R = 8.314$ J·mol$^{-1}$·K$^{-1}$) are given

TABLE 5

ERAS parameters for methanol(1) + linear polyether(2) mixtures at 298.15 K

| Linear polyether | $K_{AB}$ | $\Delta h^*_{AB}$ / kJ·mol$^{-1}$ | $\Delta v^*_{AB}$ / cm$^3$·mol$^{-1}$ | $X_{AB}$ / J·cm$^{-3}$ |
|---|---|---|---|---|
| 2,5-dioxahexane | 18 | −25 | −16 | 17 |
| 3,6-dioxaoctane | 14.8 | −25 | −19.8 | 12 |
| 2,5,8-trioxanonane | 8.5 | −25 | −16 | 14.8 |
| 2,5,8,11-tetradodecane | 8.5 | −30 | −24.2 | 17 |
| 2,5,8,11,14-pentaoxapentadecane | 8.5 | −34.2 | −27 | 22 |

$^a$ $K_{AB}$, association constant of component A with component B; $\Delta h^*_{AB}$, association enthalpy of component A with component B; $\Delta v^*_{AB}$, association volume of component A with component B; $X_{AB}$, physical parameter

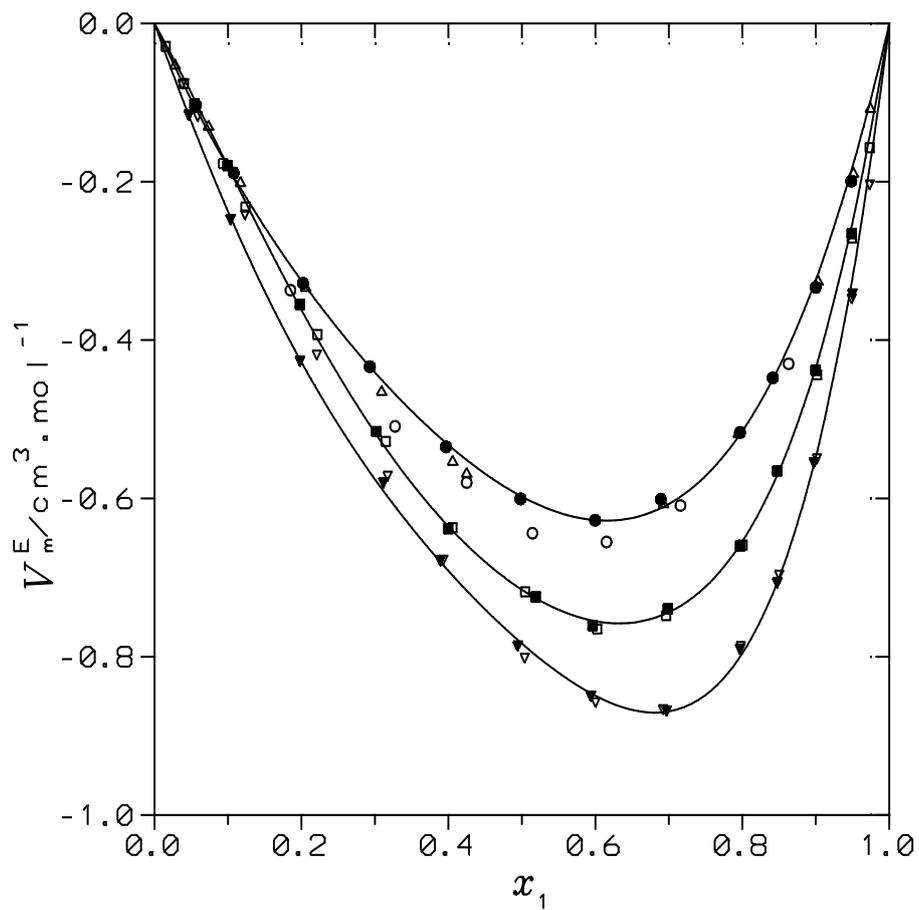

**Figure 1** $V_m^E$ at 298.15 K for methanol(1) + CH$_3$-O-(CH$_2$CH$_2$O)$_n$-CH$_3$(2) mixtures. Points, experimental data: (●) (this work); (Δ) [11]; (O) [28] $n = 2$; (■), (this work); (□) [11]; $n = 3$; (▼), (this work); ▽ [11], $n = 4$. Lines, calculations using equation (2) with coefficients from Table 3.

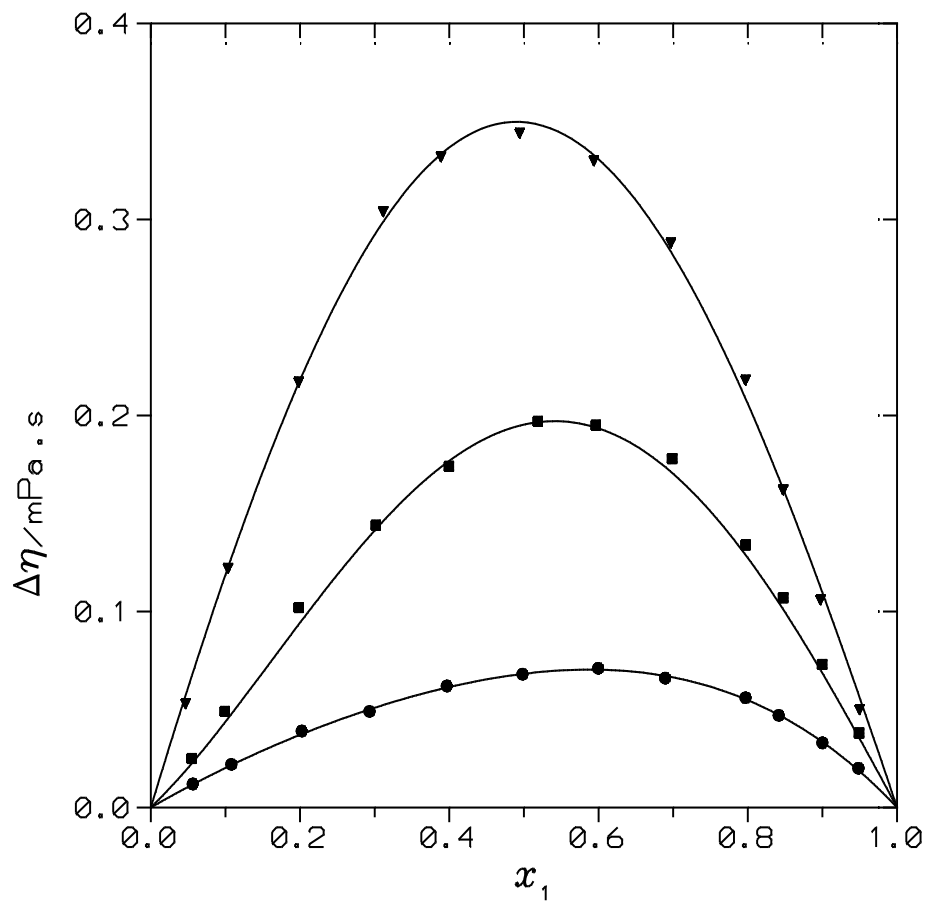

**Figure 2** $\Delta\eta$ for methanol(1) + $CH_3$-O-$(CH_2CH_2O)_n$-$CH_3$(2) systems at 298.15 K. Symbols, experimental data (this work): (●), $n = 2$; (■), $n = 3$; (▼), $n = 4$. Lines, calculations using equation (2) with coefficients from Table 3.

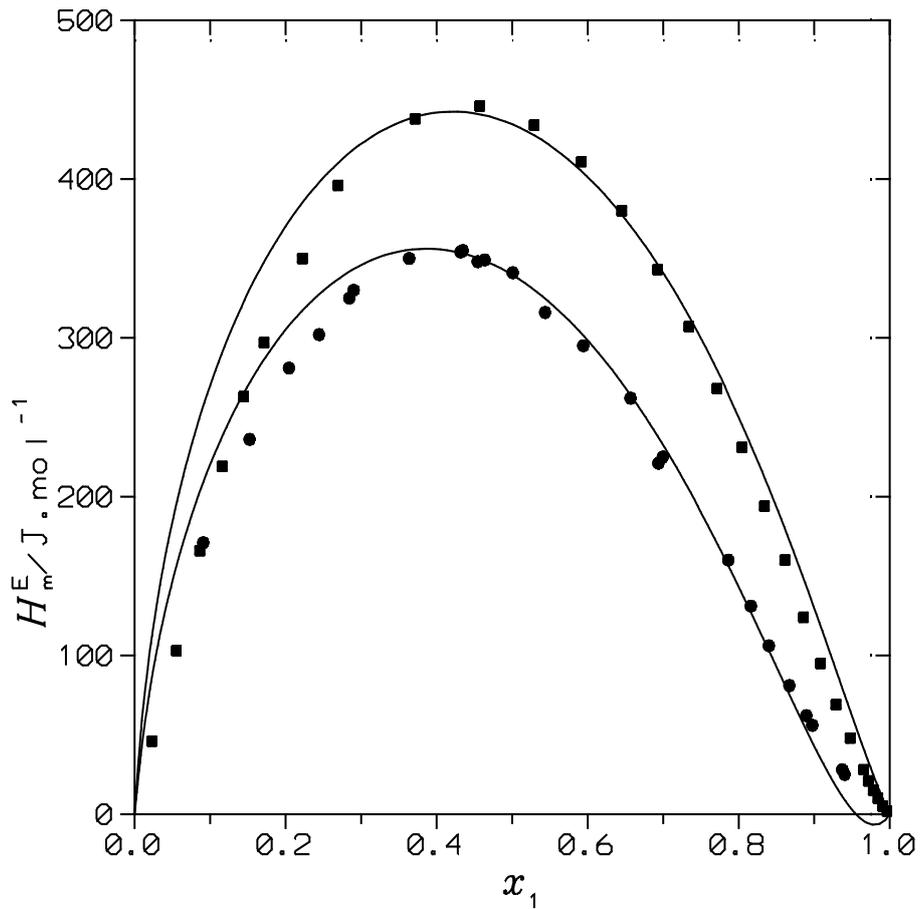

**Figure 3** $H_m^E$ at 298.15 K for methanol(1) + $CH_3$-O-$(CH_2CH_2O)_n$-$CH_3$(2) systems. Points, experimental data: (●) [15], $n = 1$; (■), $n = 2$ [54]. Lines, ERAS calculations using coefficients from Table 5.

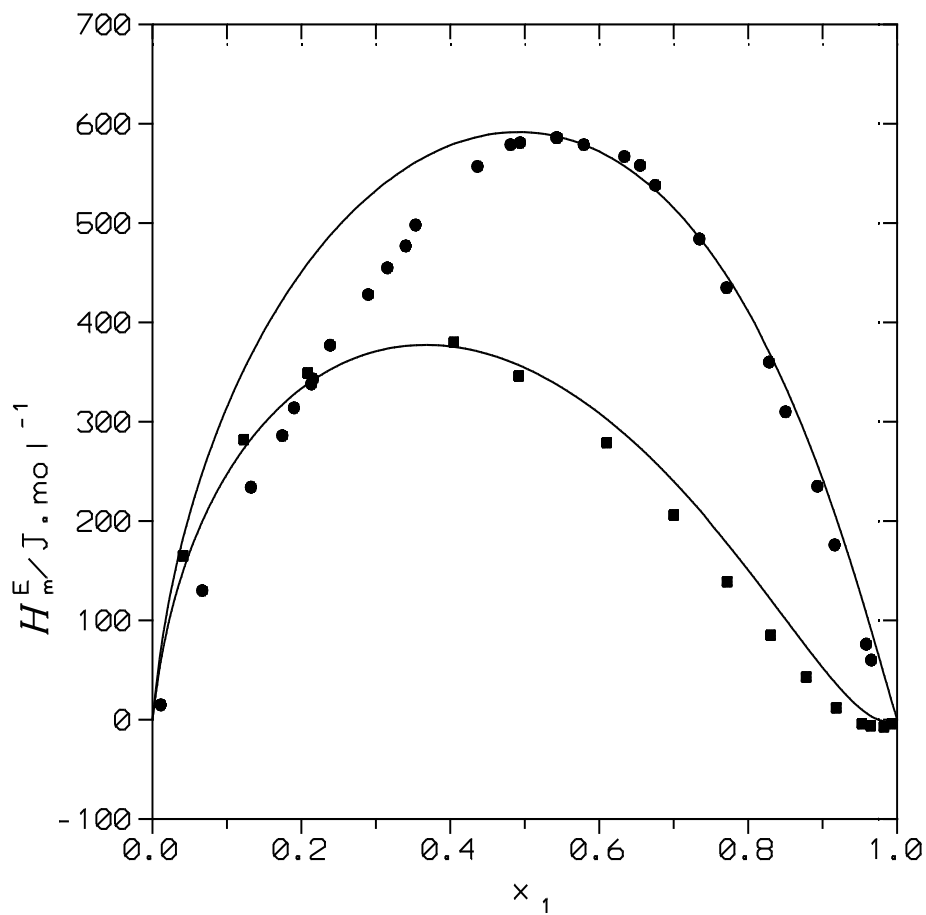

**Figure 4**   $H_m^E$ at 298.15 K for methanol(1) + linear polyether(2) systems. Points, experimental data: (●), $CH_3$-O-$(CH_2CH_2O)_4$-$CH_3$ [55]; (■), 3,6-dioxaoctane [54]. Lines, ERAS calculations using coefficients from Table 5.

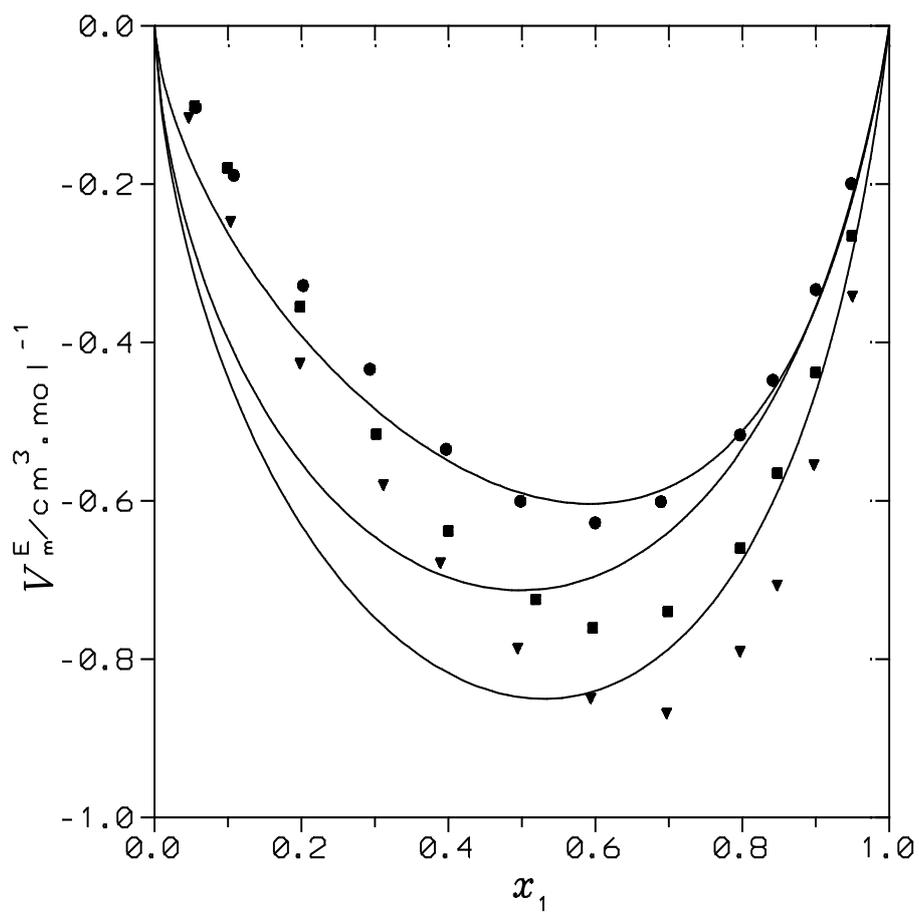

**Figure 5** $V_m^E$ at 298.15 K for methanol(1) + $CH_3$-O-$(CH_2CH_2O)_n$-$CH_3$(2) systems. Points, experimental data (this work): (●), $n = 2$; (■), $n = 3$; (▼), $n = 4$. Lines, ERAS calculations using coefficients from Table 5.

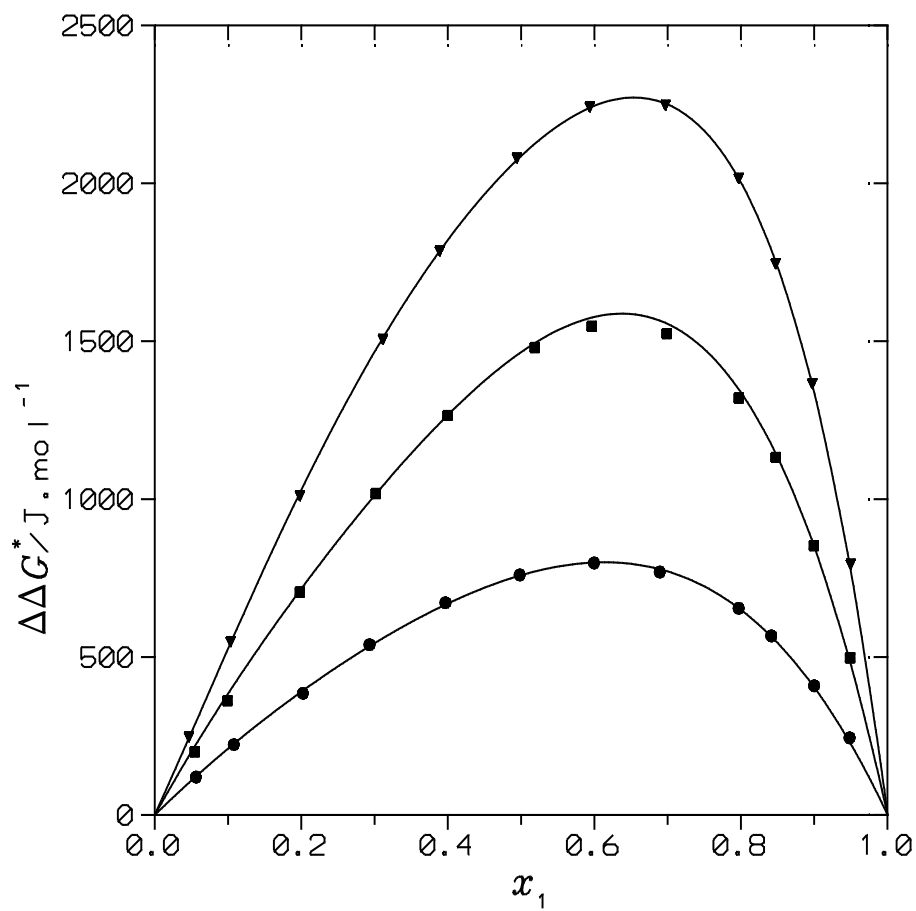

**Figure 6**  $\Delta\Delta G^*$ for methanol(1) + $CH_3$-O-$(CH_2CH_2O)_n$-$CH_3$(2) systems at 298.15 K. Symbols, experimental results (this work): (●), $n = 2$; (■), $n = 3$; (▼), $n = 4$. Lines, calculations using equation (2) with coefficients from Table 3.

SUPPLEMENTARY MATERIAL

# VOLUMETRIC AND VISCOSIMETRIC MEASUREMENTS FOR METHANOL + $CH_3$-O-$(CH_2CH_2O)_n$-$CH_3$ ($n$ = 2,3,4) MIXTURES AT (293.15-303.15) K AND ATMOSPHERIC PRESSURE. APPLICATION OF THE ERAS MODEL

**Juan A. González,\* Francisco J. Martínez, Luis F. Sanz, Fernando Hevia, Isaías García de la Fuente, José C. Cobos**

G.E.T.E.F., Departamento de Física Aplicada, Facultad de Ciencias, Universidad de Valladolid. Paseo de Belén, 7, 47011 Valladolid, Spain.

\*e-mail: jagl@termo.uva.es; Tel: +34-983-423757

1. **Viscosity correlations**

Viscosity data provided in this work have been correlated by means of the following semi-empirical equations. For dynamic viscosities the applied equations are:

Grunberg-Nissan:

$$\eta = \exp[(x_1 \ln \eta_1 + x_2 \ln \eta_2 + x_1 x_2 G_{12})] \qquad (S1)$$

Hind:

$$\eta = x_1^2 \eta_1 + x_2^2 \eta_2 + 2 x_1 x_2 \eta_{12} \qquad (S2)$$

Frenkel:

$$\ln \eta = x_1^2 \eta_1 + x_2^2 \eta_2 + 2 x_1 x_2 \ln F_{12} \qquad (S3)$$

Katti-Chaudhri:

$$\ln(\eta V) = x_1 \ln(\eta_1 V_1) + x_2 \ln(\eta_2 V_2) + x_1 x_2 \frac{W_{12}}{RT} \qquad (S4)$$

These equations have only an adjustable parameter ($G_{12}, \eta_{12}, F_{12}, W_{12}$, respectively).

For kinematic viscosities, equations with two adjustable parameters have been used:

McAllister:

$$\ln \nu = x_1^3 \ln \nu_1 + 3 x_1^2 x_2 \ln Z_{12} + 3 x_1 x_2^2 \ln Z_{21} + x_2^3 \ln \nu_2 - \ln(x_1 + x_2 \frac{M_2}{M_1}) \qquad (S5)$$

$$+ 3 x_1^2 x_2 \ln(\frac{2}{3} + \frac{M_2}{3 M_1}) + 3 x_1 x_2^2 \ln(\frac{1}{3} + \frac{2 M_2}{3 M_1}) + x_2^3 \ln(\frac{M_2}{M_1})$$

and Heric:

$$\ln \nu = x_1 \ln \nu_1 + x_2 \ln \nu_2 + x_1 \ln M_1 + x_2 \ln M_2 - \ln(x_1 M_1 + x_2 M_2) \qquad (S6)$$

$$+ x_1 x_2 [\alpha_{12} + \alpha'_{12}(x_1 - x_2)]$$

being the parameters to be fitted, $Z_{12}, Z_{21}$ and $\alpha_{12}, \alpha'_{12}$, respectively.

TABLE S1

Densities, $\rho$, excess molar volumes, $V_m^E$, dynamic viscosities, $\eta$, and deviations of viscosity from the linear dependence on mole fraction, $\Delta\eta$, for methanol(1) + polyether(2) mixtures at temperature $T$ and 0.1 MPa.

| $x_1$ | $\rho$/g.cm$^{-3}$ | $V_m^E$/cm$^3$.mol$^{-1}$ | $\eta$/mPa.s | $\Delta\eta$/mPa.s |
|---|---|---|---|---|
| \multicolumn{5}{c}{Methanol(1) + 2,5,8-trioxanonane(2); $T$/K = 293.15 K} |
| 0.0560 | 0.942327 | − 0.1049 | 1.064 | 0.012 |
| 0.1076 | 0.940435 | − 0.1900 | 1.050 | 0.023 |
| 0.2020 | 0.936398 | − 0.3273 | 1.015 | 0.041 |
| 0.2929 | 0.931629 | − 0.4334 | 0.990 | 0.054 |
| 0.3965 | 0.924931 | − 0.5333 | 0.951 | 0.066 |
| 0.4979 | 0.916514 | − 0.5964 | 0.909 | 0.073 |
| 0.5993 | 0.905508 | − 0.6235 | 0.863 | 0.077 |
| 0.6891 | 0.892496 | − 0.5976 | 0.814 | 0.072 |
| 0.7967 | 0.870965 | − 0.5127 | 0.747 | 0.058 |
| 0.8412 | 0.859143 | − 0.4447 | 0.718 | 0.050 |
| 0.8996 | 0.839967 | − 0.3285 | 0.674 | 0.035 |
| 0.9480 | 0.819736 | − 0.1979 | 0.637 | 0.022 |
| \multicolumn{5}{c}{Methanol(1) + 2,5,8-trioxanonane(2); $T$/K = 298.15} |
| 0.0560 | 0.937345 | − 0.1032 | 0.984 | 0.012 |
| 0.1076 | 0.935453 | − 0.1891 | 0.970 | 0.022 |
| 0.2020 | 0.931422 | − 0.3281 | 0.940 | 0.039 |
| 0.2929 | 0.926647 | − 0.4339 | 0.916 | 0.049 |
| 0.3965 | 0.919956 | − 0.5350 | 0.882 | 0.062 |
| 0.4978 | 0.911565 | − 0.6006 | 0.843 | 0.068 |
| 0.5993 | 0.900571 | − 0.6278 | 0.800 | 0.071 |
| 0.6891 | 0.887569 | − 0.6012 | 0.756 | 0.066 |
| 0.7967 | 0.866088 | − 0.5168 | 0.697 | 0.056 |
| 0.8411 | 0.854281 | − 0.4479 | 0.669 | 0.047 |
| 0.8996 | 0.835188 | − 0.3334 | 0.629 | 0.033 |
| 0.9480 | 0.814959 | − 0.1993 | 0.595 | 0.020 |
| \multicolumn{5}{c}{Methanol(1) + 2,5,8-trioxanonane(2); $T$/K = 303.15} |
| 0.0560 | 0.932378 | − 0.1035 | 0.903 | 0.011 |
| 0.1076 | 0.930473 | − 0.1884 | 0.890 | 0.019 |
| 0.2020 | 0.926432 | − 0.3272 | 0.863 | 0.035 |
| 0.2929 | 0.921657 | − 0.4339 | 0.840 | 0.044 |

TABLE S1 (continued)

| | | | | |
|---|---|---|---|---|
| 0.3965 | 0.914956 | − 0.5343 | 0.808 | 0.055 |
| 0.4978 | 0.906593 | − 0.6032 | 0.775 | 0.063 |
| 0.5993 | 0.895611 | − 0.6311 | 0.735 | 0.065 |
| 0.6891 | 0.882643 | − 0.6059 | 0.693 | 0.060 |
| 0.7967 | 0.861187 | − 0.5205 | 0.641 | 0.052 |
| 0.8411 | 0.849420 | − 0.4524 | 0.615 | 0.045 |
| 0.8996 | 0.830353 | − 0.3364 | 0.578 | 0.031 |
| 0.9480 | 0.810177 | − 0.2018 | 0.546 | 0.019 |
| Methanol(1) + 2,5,8,11-tetraoxadodecane(2); $T$/K = 293.15 K | | | | |
| 0.0543 | 0.984278 | − 0.1025 | 2.129 | 0.028 |
| 0.0988 | 0.982559 | − 0.1800 | 2.083 | 0.054 |
| 0.1978 | 0.978232 | − 0.3523 | 1.982 | 0.110 |
| 0.3013 | 0.972620 | − 0.5117 | 1.864 | 0.158 |
| 0.3994 | 0.965819 | − 0.6303 | 1.737 | 0.188 |
| 0.5183 | 0.954710 | − 0.7167 | 1.570 | 0.211 |
| 0.5962 | 0.945145 | − 0.7546 | 1.443 | 0.208 |
| 0.6983 | 0.927936 | − 0.7317 | 1.260 | 0.188 |
| 0.7969 | 0.903693 | − 0.6527 | 1.055 | 0.141 |
| 0.8472 | 0.886395 | − 0.5595 | 0.944 | 0.110 |
| 0.8993 | 0.863294 | − 0.4330 | 0.825 | 0.074 |
| 0.9489 | 0.833866 | − 0.2622 | 0.711 | 0.040 |
| Methanol(1) + 2,5,8,11-tetroaxadodecane(2); $T$/K = 298.15 K | | | | |
| 0.0543 | 0.979508 | − 0.1017 | 1.922 | 0.025 |
| 0.0988 | 0.977785 | − 0.1795 | 1.882 | 0.049 |
| 0.1978 | 0.973463 | − 0.3549 | 1.795 | 0.102 |
| 0.3013 | 0.967843 | − 0.5157 | 1.689 | 0.144 |
| 0.3994 | 0.961057 | − 0.6381 | 1.580 | 0.174 |
| 0.5183 | 0.949935 | − 0.7245 | 1.433 | 0.197 |
| 0.5962 | 0.940344 | − 0.7606 | 1.321 | 0.195 |
| 0.6983 | 0.923153 | − 0.7397 | 1.158 | 0.178 |
| 0.7969 | 0.898910 | − 0.6599 | 0.975 | 0.134 |
| 0.8472 | 0.881607 | − 0.5651 | 0.875 | 0.107 |
| 0.8993 | 0.858526 | − 0.4380 | 0.767 | 0.073 |
| 0.9489 | 0.829121 | − 0.2654 | 0.662 | 0.038 |

Methanol(1) + 2,5,8,11-tetraoxadodecane(2); $T$/K = 303.15 K

| | | | | |
|---|---|---|---|---|
| 0.0543 | 0.974750 | − 0.1019 | 1.733 | 0.025 |
| 0.0988 | 0.973032 | − 0.1817 | 1.696 | 0.044 |
| 0.1978 | 0.968727 | − 0.3625 | 1.621 | 0.095 |
| 0.3013 | 0.963071 | − 0.5202 | 1.527 | 0.132 |
| 0.3994 | 0.956308 | − 0.6478 | 1.434 | 0.164 |
| 0.5183 | 0.945165 | − 0.7335 | 1.302 | 0.184 |
| 0.5962 | 0.935540 | − 0.7669 | 1.203 | 0.184 |
| 0.6983 | 0.918364 | − 0.7482 | 1.058 | 0.169 |
| 0.7969 | 0.894125 | − 0.6682 | 0.894 | 0.130 |
| 0.8472 | 0.876812 | − 0.5715 | 0.804 | 0.104 |
| 0.8993 | 0.853748 | − 0.4437 | 0.705 | 0.071 |
| 0.9489 | 0.824370 | − 0.2697 | 0.608 | 0.038 |

Methanol(1) + 2,5,8,11,14-pentaoxapentadecane(2); $T$/K = 293.15 K

| | | | | |
|---|---|---|---|---|
| 0.0463 | 1.009936 | − 0.1123 | 3.726 | 0.058 |
| 0.1031 | 1.008010 | − 0.2428 | 3.616 | 0.131 |
| 0.1978 | 1.004109 | − 0.4232 | 3.408 | 0.229 |
| 0.3111 | 0.997982 | − 0.5766 | 3.132 | 0.319 |
| 0.3886 | 0.992773 | − 0.6740 | 2.919 | 0.356 |
| 0.4940 | 0.983710 | − 0.7807 | 2.592 | 0.369 |
| 0.5935 | 0.972005 | − 0.8430 | 2.250 | 0.348 |
| 0.6966 | 0.954669 | − 0.8581 | 1.868 | 0.298 |
| 0.7966 | 0.928658 | − 0.7800 | 1.467 | 0.221 |
| 0.8472 | 0.909651 | − 0.6981 | 1.243 | 0.161 |
| 0.8972 | 0.884032 | − 0.5460 | 1.024 | 0.103 |
| 0.9495 | 0.846553 | − 0.3371 | 0.803 | 0.050 |

Methanol(1) + 2,5,8,11,14-pentaoxapentadecane(2); $T$/K = 298.15 K

| | | | | |
|---|---|---|---|---|
| 0.0463 | 1.005320 | − 0.1162 | 3.300 | 0.053 |
| 0.1031 | 1.003388 | − 0.2476 | 3.209 | 0.122 |
| 0.1978 | 0.999465 | − 0.4267 | 3.036 | 0.217 |
| 0.3111 | 0.993317 | − 0.5802 | 2.803 | 0.304 |
| 0.3886 | 0.988100 | − 0.6788 | 2.611 | 0.332 |
| 0.4940 | 0.979020 | − 0.7865 | 2.326 | 0.344 |
| 0.5935 | 0.967301 | − 0.8500 | 2.031 | 0.330 |
| 0.6966 | 0.949975 | − 0.8685 | 1.696 | 0.288 |

TABLE S1 (continued)

| | | | | |
|---|---|---|---|---|
| 0.7966 | 0.923955 | − 0.7905 | 1.344 | 0.218 |
| 0.8472 | 0.904927 | − 0.7068 | 1.145 | 0.162 |
| 0.8972 | 0.879332 | − 0.5550 | 0.948 | 0.106 |
| 0.9495 | 0.841832 | − 0.3419 | 0.744 | 0.050 |
| Methanol(1) + 2,5,8,11,14-pentaoxapentadecane(2); $T$/K = 303.15 K | | | | |
| 0.0463 | 1.000703 | − 0.1164 | 2.916 | 0.046 |
| 0.1031 | 0.998764 | − 0.2485 | 2.845 | 0.116 |
| 0.1978 | 0.994828 | − 0.4289 | 2.691 | 0.197 |
| 0.3111 | 0.988666 | − 0.5838 | 2.493 | 0.280 |
| 0.3886 | 0.983427 | − 0.6819 | 2.334 | 0.312 |
| 0.4940 | 0.974340 | − 0.7925 | 2.089 | 0.329 |
| 0.5935 | 0.962606 | − 0.8575 | 1.829 | 0.316 |
| 0.6966 | 0.945281 | − 0.8790 | 1.537 | 0.279 |
| 0.7966 | 0.919251 | − 0.8016 | 1.225 | 0.215 |
| 0.8472 | 0.900208 | − 0.7167 | 1.047 | 0.163 |
| 0.8972 | 0.874625 | − 0.5647 | 0.871 | 0.111 |
| 0.9495 | 0.837107 | − 0.3479 | 0.683 | 0.052 |

[a] the uncertainties, $u$, are: $u(T) = \pm\, 0.01$ K; $u(x_1) = \pm\, 0.0001$; $u(\rho) = \pm\, 5 \cdot 10^{-5}$ g·cm$^{-3}$; $u(V_m^E) = \pm\, (0.01\, |V_{m,min}^E| + 0.005$ cm$^3$·mol$^{-1}$); $u(\eta) = \pm\, 0.01 \cdot \eta$ mPa·s and $u(\Delta\eta) = \pm\, 0.02 \cdot \Delta\eta$ mPa·s

TABLE S2

Parameters of activation of viscous flow: enthalpy, $\Delta H^*$, entropy, $\Delta S^*$, Gibbs energy, $\Delta G^*$, and the deviation of $\Delta G^*$ from the linear dependence on mole fraction for methanol(1) + linear polyether(2) mixtures mixtures at 298.15 K.

| $x_1$ | $\Delta H^*$/J.mol$^{-1}$ | $\Delta S^*$/J.mol$^{-1}$.K$^{-1}$ | $\Delta G^*$/J.mol$^{-1}$ | $\Delta(\Delta G^*)$/J.mol$^{-1}$ |
|---|---|---|---|---|
| | Methanol(1) + 2,5,8-trioxanonane(2) | | | |
| 0 | 11362 | − 10.7 | 14563 | 0 |
| 0.0560 | 11326 | − 10.4 | 14426 | 120 |
| 0.1076 | 11354 | − 9.8 | 14292 | 223 |
| 0.2020 | 11231 | − 9.4 | 14021 | 385 |
| 0.2929 | 11316 | − 8.2 | 13757 | 540 |
| 0.3965 | 11191 | − 7.5 | 13414 | 672 |
| 0.4978 | 10999 | − 6.8 | 13036 | 759 |
| 0.5993 | 10987 | − 5.4 | 12608 | 797 |
| 0.6891 | 11040 | − 3.8 | 12167 | 769 |
| 0.7967 | 10506 | − 3.5 | 11558 | 654 |
| 0.8411 | 10526 | − 2.5 | 11267 | 567 |
| 0.8996 | 10455 | − 1.3 | 10840 | 409 |
| 0.9480 | 10534 | 0.3 | 10453 | 244 |
| 1 | 10495 | 1.8 | 9971 | 0 |
| | Methanol(1) + 2,5,8,11-tetraoxadodecane(2) | | | |
| 0 | 14631 | − 7.5 | 16852 | 0 |
| 0.0543 | 14471 | − 7.4 | 16679 | 200 |
| 0.0988 | 14475 | − 6.9 | 16534 | 362 |
| 0.1978 | 14129 | − 6.9 | 16197 | 706 |
| 0.3013 | 14021 | − 6.0 | 15797 | 1019 |
| 0.3994 | 13448 | − 6.4 | 15369 | 1265 |
| 0.5183 | 13084 | − 5.6 | 14766 | 1480 |
| 0.5962 | 12681 | − 5.4 | 14296 | 1547 |
| 0.6983 | 12141 | − 4.8 | 13571 | 1524 |
| 0.7969 | 11454 | − 4.1 | 12688 | 1320. |
| 0.8472 | 11102 | − 3.5 | 12154 | 1132 |
| 0.8993 | 10827 | − 2.3 | 11517 | 854 |
| 0.9489 | 10724 | − 0.3 | 10820 | 498 |
| 1 | 10495 | 1.8 | 9971 | 0 |

TABLE S2 (continued)

Methanol(1) + 2,5,8,11,14-pentaoxapentadecane(2)

| | | | | |
|---|---|---|---|---|
| 0 | 17502 | − 3.9 | 18669 | 0 |
| 0.0463 | 17419 | − 3.7 | 18513 | 247 |
| 0.1031 | 17036 | − 4.3 | 18321 | 549 |
| 0.1978 | 16766 | − 4.0 | 17960 | 1011 |
| 0.3111 | 16152 | − 4.4 | 17469 | 1506 |
| 0.3886 | 15835 | − 4.2 | 17075 | 1786 |
| 0.4940 | 15242 | − 4.1 | 16453 | 2080 |
| 0.5935 | 14587 | − 3.9 | 15748 | 2242 |
| 0.6966 | 13649 | − 4.1 | 14857 | 2247 |
| 0.7966 | 12583 | − 3.9 | 13756 | 2016 |
| 0.8472 | 11924 | − 3.8 | 13046 | 1747 |
| 0.8972 | 11173 | − 3.6 | 12230 | 1366 |
| 0.9495 | 11152 | − 0.2 | 11204 | 794 |
| 1 | 10495 | 1.8 | 9971 | 0 |